\documentclass[conference]{IEEEtran}
\IEEEoverridecommandlockouts

\usepackage{cite}
\usepackage{subfig}
\usepackage{amsmath,amssymb,amsfonts}
\usepackage{algorithmic}
\usepackage{graphicx}
\usepackage{textcomp}
\usepackage{xcolor}
\usepackage{siunitx}
\usepackage{longtable}
\def\BibTeX{{\rm B\kern-.05em{\sc i\kern-.025em b}\kern-.08em
    T\kern-.1667em\lower.7ex\hbox{E}\kern-.125emX}}
\begin{document}

\title{2T1R Regulated Memristor Conductance Control Array Architecture for Neuromorphic Computing using 28nm CMOS Technology\\

\thanks{This research work is funded by Bundesministerium für Bildung und Forschung (BMBF grant number 16ME0398K), under the NEUOROTEC II project of Forschungszentrum, J{\"u}lich, Germany}
}

\author{
    \IEEEauthorblockN{
        Neethu Kuriakose\IEEEauthorrefmark{1},
        Arun Ashok\IEEEauthorrefmark{1},
        Christian Grewing\IEEEauthorrefmark{1},\\
        André Zambanini\IEEEauthorrefmark{1}, and
        Stefan van Waasen\IEEEauthorrefmark{1}\IEEEauthorrefmark{2}
   }\\
    \IEEEauthorblockA{
        \IEEEauthorrefmark{1}Central Institute of Engineering,
        Electronics and Analytics -- Electronic Systems (ZEA-2),\\
        Forschungszentrum Jülich GmbH,
        52425 Jülich, Germany\\
        \IEEEauthorrefmark{2}Faculty of Engineering, 
        Communication Systems, 
        University of Duisburg-Essen, 
        47057 Duisburg, Germany\\
        Email: n.kuriakose@fz-juelich.de
    }
}

\maketitle
\begin{abstract}

Memristors are promising devices for scalable and low power, in-memory computing to improve the energy efficiency of a rising computational demand. The crossbar array architecture with memristors is used for vector matrix multiplication (VMM) and acts as kernels in neuromorphic computing. The analog conductance control in a memristor is achieved by applying voltage or current through it. A basic 1T1R array is suitable to avoid sneak path issues but suffer from wire resistances, which affects the read  and write procedures. A conductance control scheme with a regulated voltage source will improve the architecture and reduce the possible potential divider effects. A change in conductance is also possible with the provision of a regulated current source and measuring the voltage across the memristors. A regulated 2T1R memristor conductance control architecture is proposed in this work, which avoids the potential divider effect and virtual ground scenario in a regular crossbar scheme, as well as conductance control by passing a regulated current through memristors. The sneak path current is not allowed to pass by the provision of ground potential to both terminals of memristors.
\end{abstract}


\section{Introduction}

Computing in memory (CIM) is a new computing paradigm in which the computation and storage are done in situ\cite{b1}. Emerging non-volatile memory~(NVM) devices aid in the development of the CIM architecture, where the non-volatile conductance is utilized to store data. 
NVM devices arranged in a crossbar array pattern emulate vector matrix multiplication~(VMM) operation and show high parallelism. The conductance state of the device is continued even after the stimulus is removed. A memristor is a promising NVM device for computing in memory due to higher memory densities, faster read/write capability, and lower power dissipation\cite{b2}. However, sneak path current problems exist, causing cross-talk interference between adjacent memristor cells, which may result in misinterpretation of the crossbar readout. The sneak path issue could be tackled by using a selector device, which is commonly called 1S1R structure \cite{b3}\cite{b4}. If the selector device is a transistor, the structure is called 1T1R (or 2T1R structure if the selection transistor is given on both terminals of the memristor, for bipolar biasing) structure. IR drops exist due to the line resistance connecting the devices in array architecture, which creates voltage drop and affects the required conductance value, if the memristor is in read or write mode \cite{b5}. Moreover, a virtual ground concept prevails at one end of the memristor as shown in Fig. 1, if a transimpedence amplifier (TIA) is used for read process.
\begin{figure}[h]
    \centering
   \includegraphics[width=1\linewidth]{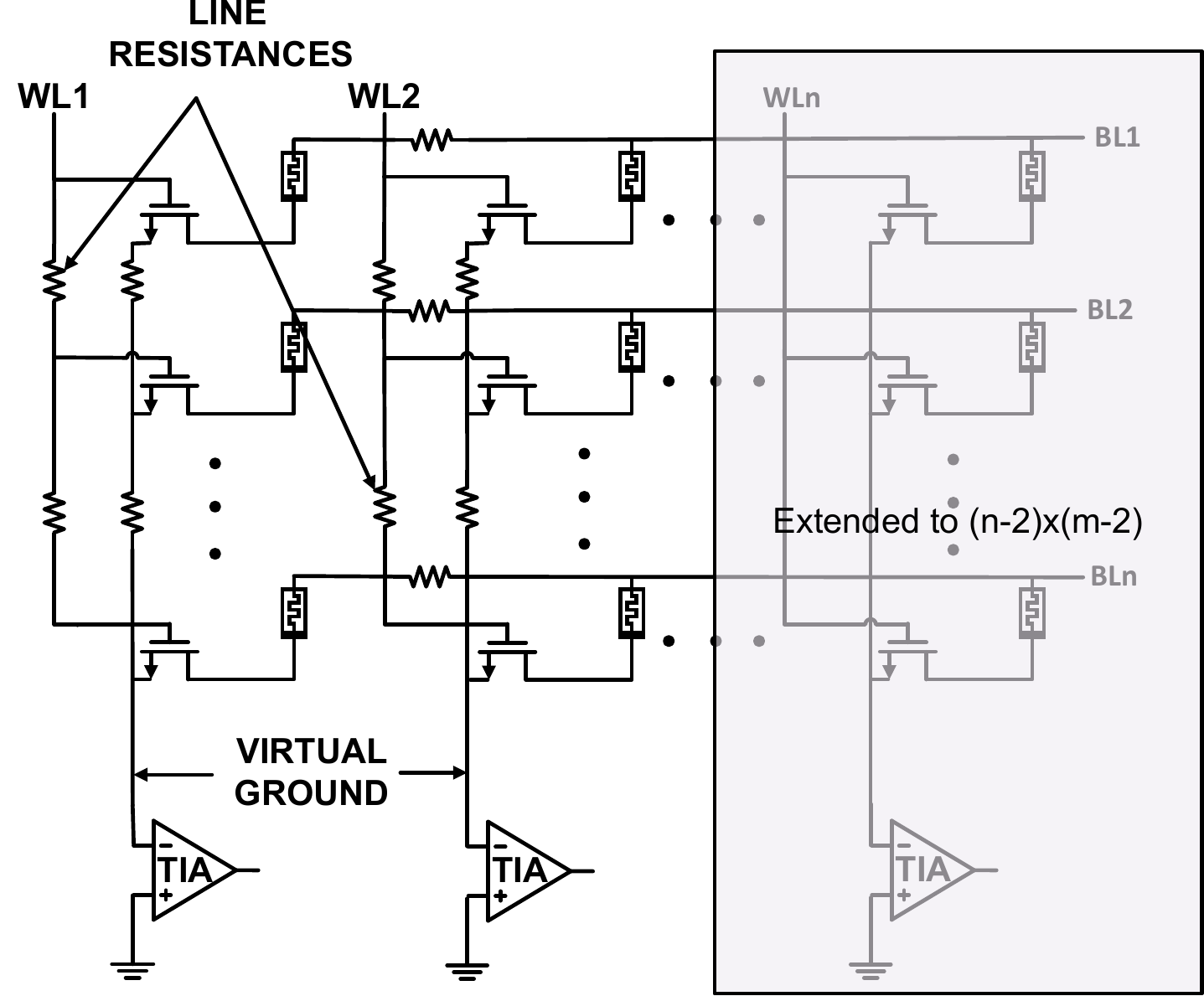}
    \caption{1T1R including line resistances and concept of virtual ground}
\end{figure}

In this article, we propose a regulated read-write control to a memristor, which overcomes the mentioned effects of potential differences and virtual ground by providing local regulation and ensuring local ground. Two modes of memristor control are introduced, where the effect of line resistance and conductance can be controlled. In the proposed system, the current is read back with a current mode SAR analog to digital converter (ADC) in which, the current is sourced by the ADC to the array architecture rather than the approach using a TIA. This results in a reduced overall power consumption of the system. The sneak path current issue is resolved by grounding both terminals of the memristor to ground by analog switches.

Section II provides an insight into the JART VCM v1b memristor model, which is used in our simulations. Section III explains two different conductance schemes. First, the voltage mode, in which a regulated voltage is applied to either of the memristor electrodes, keeping the opposite side grounded, and next the current mode, in which a regulated current is allowed to pass through a memristor, which sets the memristor from a high resistance state (HRS) to a low resistance state (LRS). Section IV describes a 2T1R array incorporating both modes of operation to switch between read and write, for an array size of 2x2.

\section{ReRAM Device Model}
Resistive switching cells based on valence change mechanism (VCM) are foreseen to be used in computation in memory and neuromorphic applications because of their non-linearity, multilevel behavior, and switching statistics. VCM devices consist of a metal-metal oxide-metal structure\cite{b6} as shown in Fig. 2a. The work function of the two metal electrodes is different. The metal electrode with higher and lower work functions corresponds to the active electrode (AE) or bottom electrode and the ohmic electrode (OE) or top electrode respectively. The V-I characteristics, of the memristor model, with a sweep rate of \SI{0.67}{\volt/\sec} is depicted in Fig. 2b  

The oxide layer is highly insulating after the fabrication, therefore an electroforming step is performed initially, applying a high voltage with which oxygen vacancies are created thus reducing the resistance of the device. The concentration of oxygen vacancies is increased during the SET process and decreased using the RESET process. The SET occurs when a negative potential is applied to the active electrode and the device enters the LRS state. A RESET occurs when a positive potential is applied to the active electrode and the device enters the HRS state. There exist device-to-device and cycle-to-cycle variability involved in resistive switching phenomena.

The model used in simulations is the JART VCM v1b model, developed for HfO2/TiOx based ReRAM cell \cite{b7}. The model is implemented using VerilogA and with associated circuitry, Cadence is used for circuit simulations. The model parameters are given in Table I. 

\begin{table}[h]
\caption{Deterministic model parameters of JART VCM V1b}
\begin{center}
\begin{tabular}{|c|c|}
\hline
\cline{1-2} 
\textbf{\textit{Model Parameters}}& \textbf{\textit{Value}} \\
\hline

$\textit{A}_\text{det}$ & \SI{6.36d-15}{\meter^{2}}\\
\hline
$\textit{N}_\text{disc}$ & \SI{0.00826}{\meter^{-3}} \\
\hline
$\textit{l}_\text{cell}$ & \SI{3}{\nano\meter}\\
\hline
$\textit{l}_\text{plug}$ & \SI{2.6}{\nano\meter}\\
\hline
$\textit{R}_\text{series}$ & \SI{1.37}{\kilo\Omega} \\
\hline
$\textit{R}_\text{line}$ & \SI{719}{\Omega}\\
\hline
$\textit{R}_\text{TiOx}$ & \SI{650}{\Omega} \\
\hline
$\gamma_{0}$ & \SI{2d13}{\hertz} \\
\hline
$\textit{R}_\text{th0,SET}$ & \SI{15.72d6}{\kelvin/\watt}\\
\hline
$\textit{e}\phi_\text{Bn0}$ & \SI{0.18}{e\volt} \\
\hline
$\textit{R}_\text{th,line}$ & \SI{90471.47}{\kelvin/\watt}\\
\hline
$\alpha_\text{line}$ & \SI{3.92d-3}{1/\kelvin} \\
\hline
$\textit{A}^{*}$ & \SI{6.01d5}{\ampere/\meter^{2}\kelvin^{2}}\\
\hline
$\textit{kb}$ & \SI{1.38d-23}{\joule/\kelvin}\\
\hline
$\textit{rd}$ & \SI{45}{\nano\meter} \\
\hline
$\textit{a}$ & \SI{0.25}{\nano\meter} \\
\hline
$\textit{l}_{det}$ & \SI{0.4}{\nano\meter}\\
\hline
$\textit{T}_0 $ & \SI{293}{\kelvin} \\
\hline

\end{tabular}
\label{tab1}
\end{center}
\end{table}

\begin{figure}[h]
    \centering
   \subfloat[ ] {\includegraphics[width=0.45\linewidth]{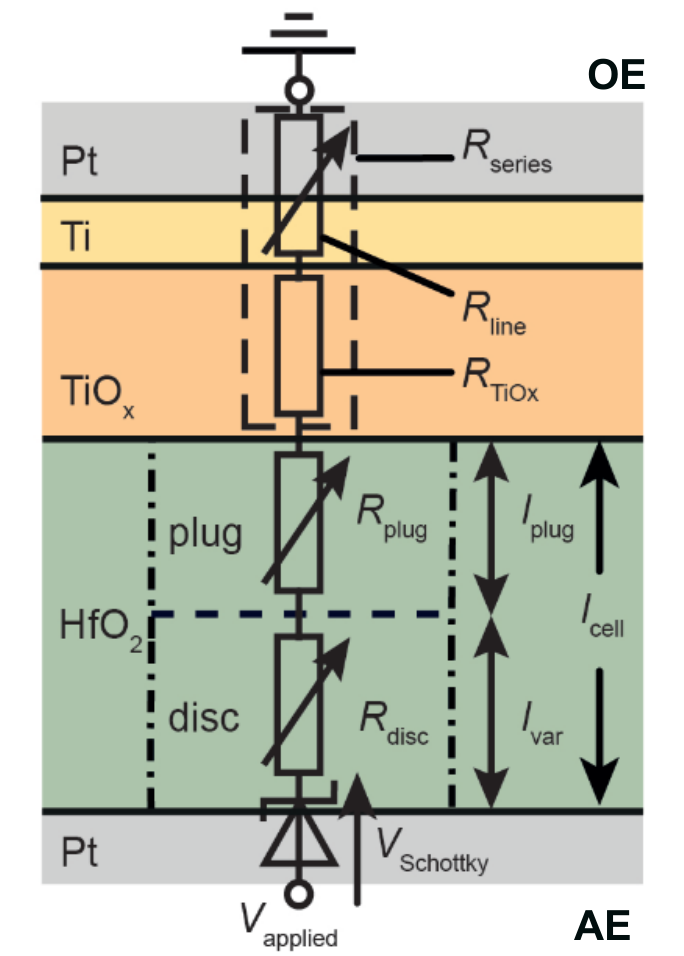}}\hfill
   \subfloat[ ] {\includegraphics[width=0.55\linewidth]{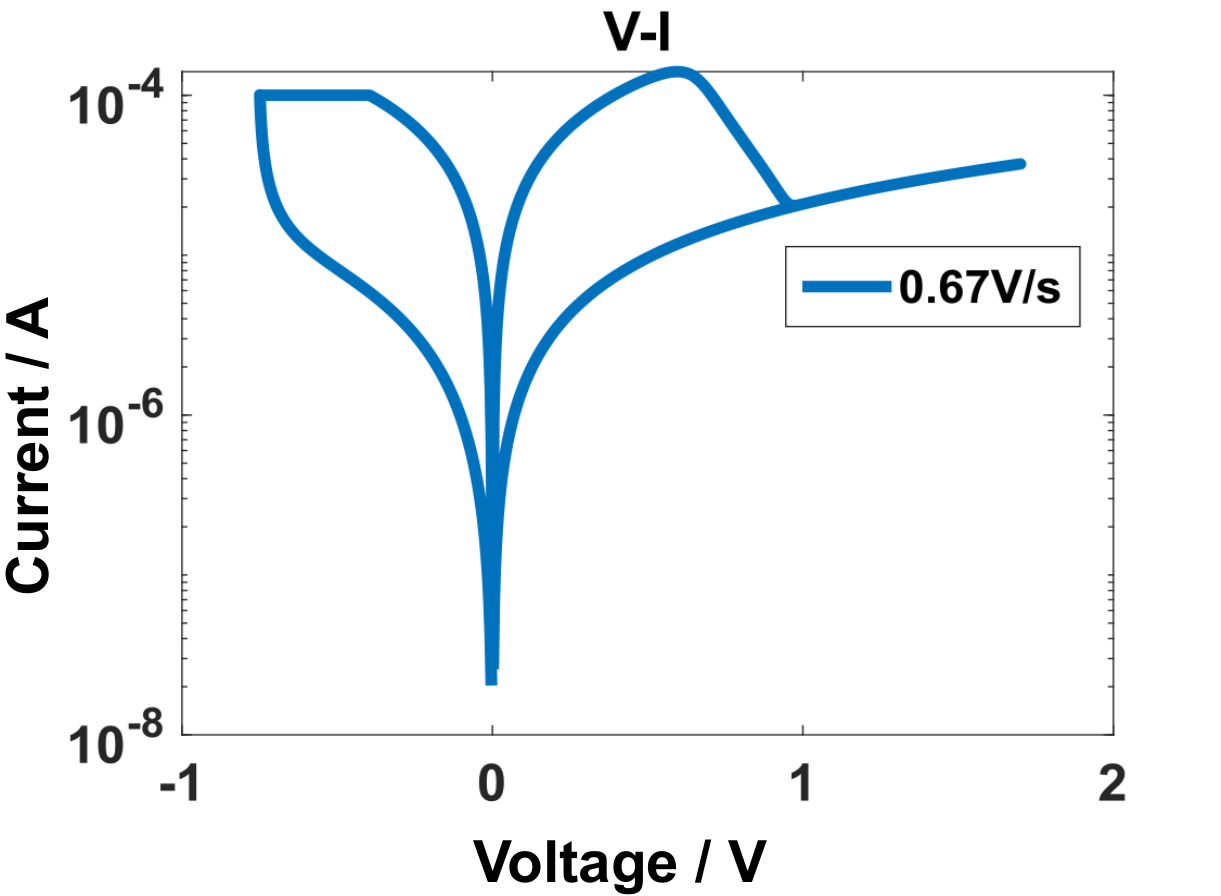}}
    \caption{Equivalent circuit diagram for the electrical model (a) of the Pt/HfO2/TiOx/Pt device \cite{b4}. The voltage is applied to the bottom electrode. The V-I characteristics correspond to the voltage sweep applied to the model (b)}
\end{figure}

\section{Memristor Conductance Control Modes}
\subsection{Overview}
The conductance or resistance of a memristor can be controlled by either controlling the voltage across it or by controlling the current through it. The pulses or sweep rates in which these electrical control techniques are applied have a large impact on the memristor conductivity. The two significant states, HRS and LRS correspond to two distinguished conductance states. However, an analog conductance change could be achieved by proper control of conductance control signals. Two different modes are defined for conductance control namely, voltage mode and current mode.

\begin{figure}[h]
    \centering
   \includegraphics[width=\linewidth]{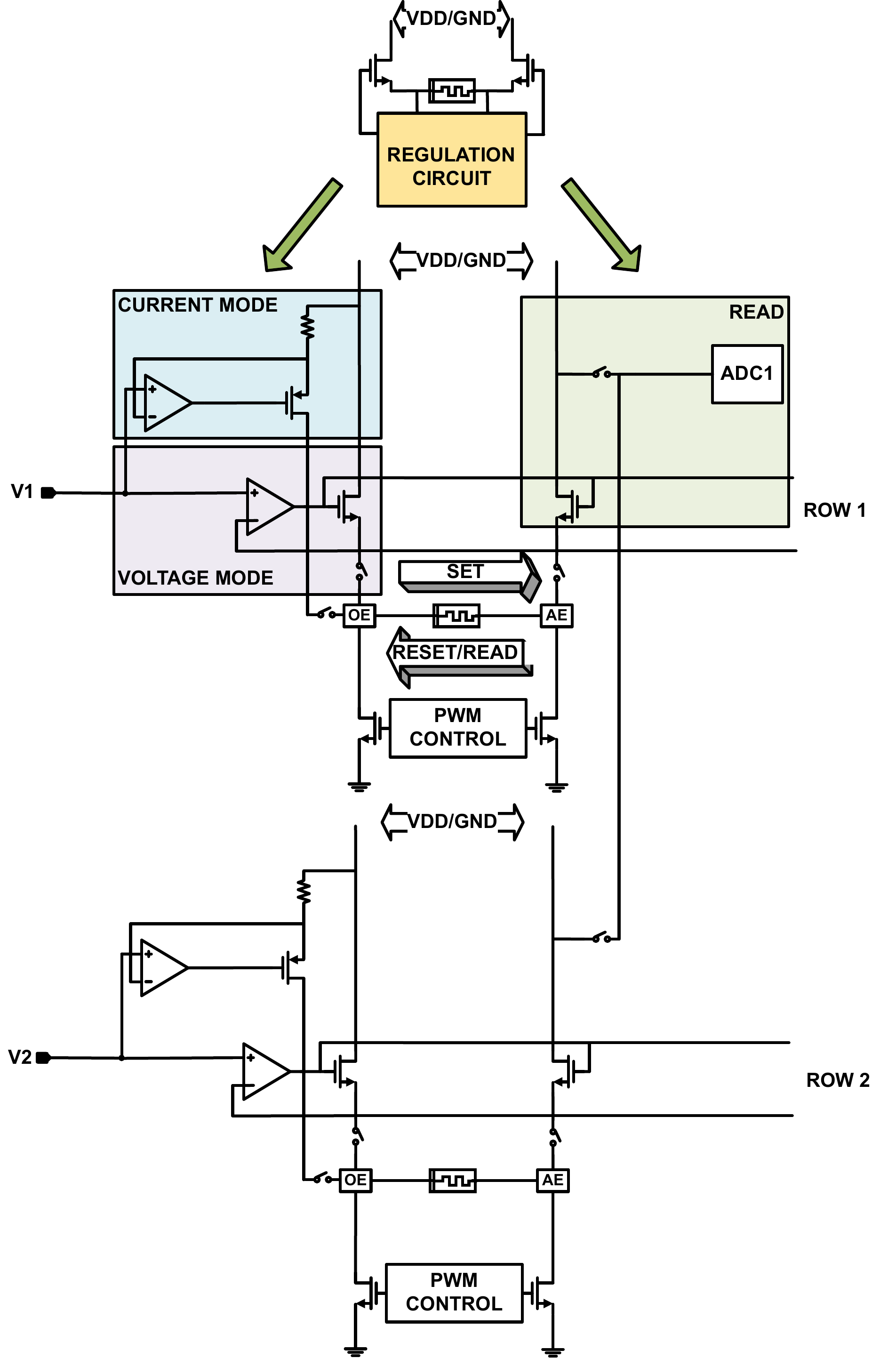}
    \caption{2T1R regulated architecture for 2x1 structure}
\end{figure}

In both modes, a regulated architecture with wide swing operational amplifiers\cite{b8} is utilized to provide a regulated signal towards the memristor ends for read and write operations as shown in Fig. 3. The input voltage to the operational amplifier (opamp) is provided from a digital-to-analog converter (DAC). The SET and RESET of memristors correspond to the write operation, and the application of a low voltage to read back the memristor state without changing the memristor conductance state corresponds to the read operation. During a read operation, the current mode SAR ADC acts as the supply source for regulation and the current is sourced to the memristor. The pulse width is controlled by a pulse width modulator (PWM). One opamp is shared per row and one ADC per column. Similarly, power supplies are also shared per column. The selection of devices and mode of operation is determined by switches. When not selected, the OE and AE terminals of the memristor are grounded, thus avoiding sneak path currents if any.

\subsection{Voltage Mode Operation}
In this mode, the memristor is applied with a regulated voltage pulse across them, as shown in Fig. 4(c). Asymmetrical voltage pulses are applied to the active electrode of the memristor to achieve the HRS and LRS states as in Table II. To abstain from the application of negative voltage to the active electrode of the memristor for SET operation, a voltage of \SI{-750}{\milli\volt} could be applied at the ohmic electrode, keeping the active electrode grounded, realising the same. 

\begin{figure}[h]
    \centering
     \subfloat[Current control]{\includegraphics[width=0.5\linewidth]{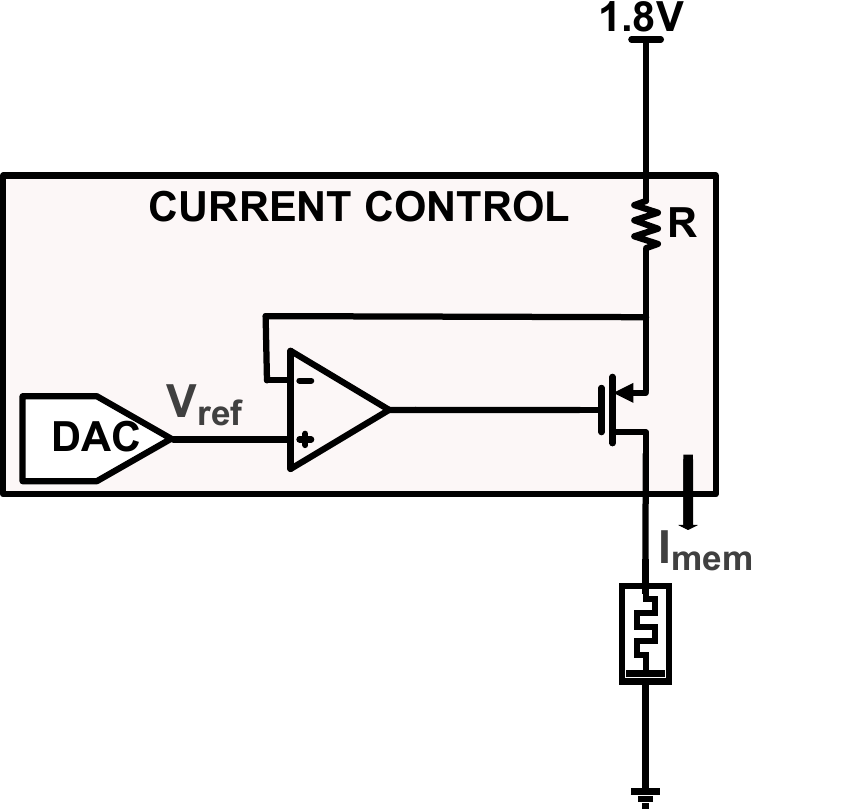}}
      \subfloat[Read control ]{\includegraphics[width=0.5\linewidth]{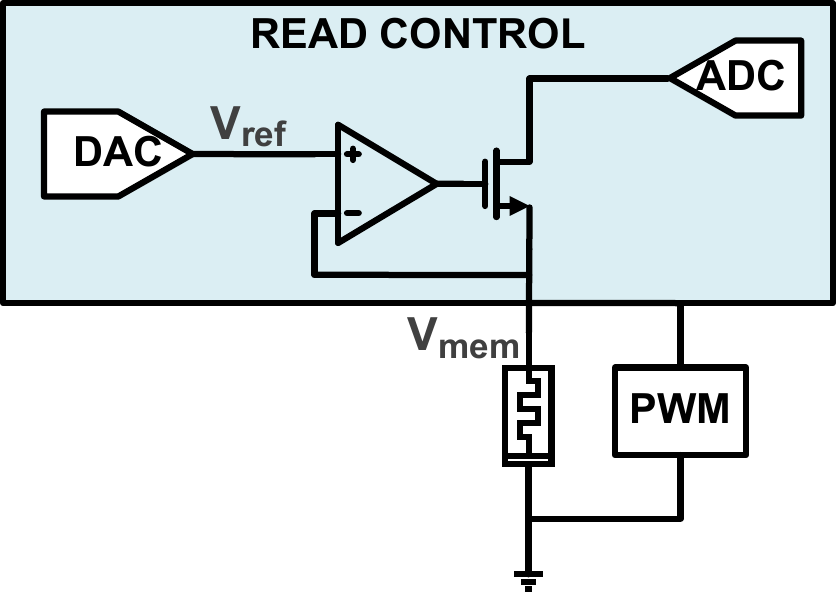}}\\
    \subfloat[Write control]{\includegraphics[width=0.5\linewidth]{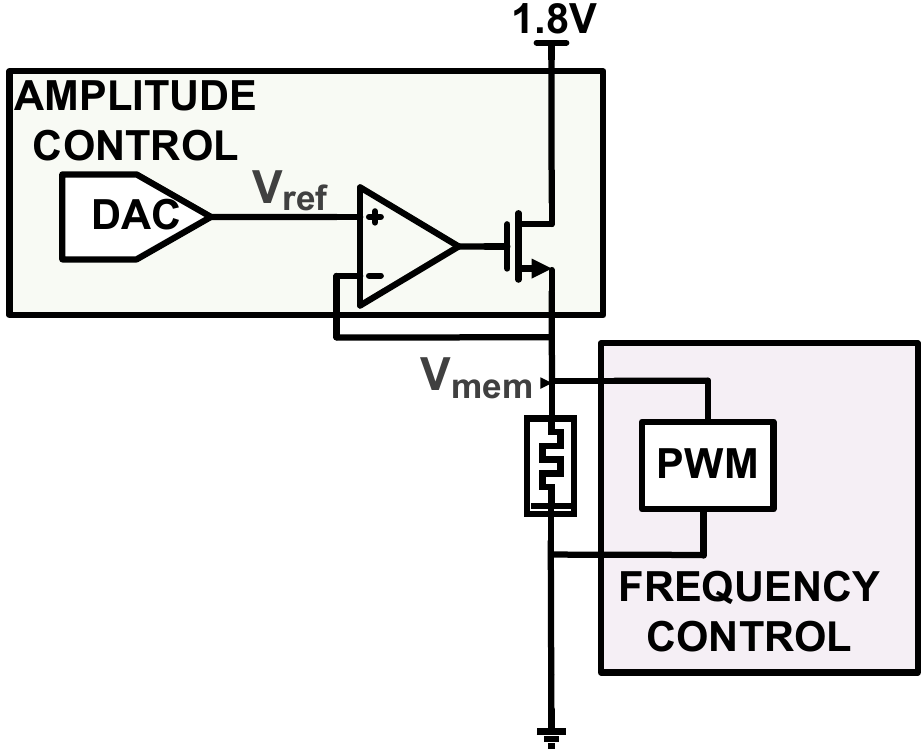}}

    \caption{Simplified view of regulator circuit for read-write operations}
\end{figure}

\begin{table}[htbp]
\caption{SET-RESET Parameters}
\begin{center}
\begin{tabular}{|c|c|c|}
\hline
\multicolumn{3}{|c|}{\textbf{Control parameters}} \\
\cline{1-3} 
\textbf{\textit{Memristor state}}& \textbf{\textit{parameter}}& \textbf{\textit{Pulse width}}  \\
\hline
HRS& \SI{1.5}{\volt}& \SI{1}{\milli\second}  \\
\hline
LRS& \SI{-750}{\milli\volt}& \SI{1}{\milli\second}\\
\hline
Read& \SI{250}{\milli\volt}& \SI{1}{\milli\second}\\
\hline
\end{tabular}
\label{tab1}
\end{center}
\end{table}
Switching between the active and the ohmic electrode is achieved with multiplexer (MUX) selector circuits, which results in a 2T1R configuration to provide regulated voltages on either memristor end, keeping the other end grounded.

The 8-bit DAC provides reference voltage $\textit{V}_\text{ref} $ to opamp configuration. The R-2R DAC is designed for \SI{6.25}{\milli\volt}  to \SI{1.6}{\volt} with step size of \SI{6.25}{\milli\volt}. The opamp is used as a voltage comparator, which compares the reference voltage and the voltage at the memristor electrode. An NMOS is utilised as a pass element in a common drain configuration. This transistor acts as the current compliance device for the memristor connected to it. The voltage at the memristor end is regulated to the reference voltage of the opamp as:

\begin{equation}
    V_\text{ref} = V_\text{mem}
\end{equation}
The folded cascode opamp works with a wide input voltage ranging from \SI{0.2}{\volt} to \SI{1.6}{\volt}.

From Table II, the LRS is obtained if the reference voltage provided by DAC is \SI{750}{\milli\volt}, which will be regulated by the regulator circuit at the memristor ohmic electrode. For HRS the reference voltage should be \SI{1.5}{\volt}. However, due to the dropout condition, the maximum voltage that will be regulated by the circuit is \SI{1.2}{\volt}. In this case, the op-amp is utilised in a voltage follower configuration with MUX selectors in between.

The transient response of the above memristor model in voltage and current domain with read voltage followed by HRS and LRS configuration, including the parasitic effects, is shown in Fig. 5. The W/L ratio of the transistor is \SI{9.6}{\micro\meter/}\SI{150}{\nano\meter}. The 8-bit DAC values provide the value closer to the requirements as given in Table II. The read verify algorithm is utilized for conductance control.
\begin{figure}[htbp]
    \centering
    \includegraphics[width=1\linewidth]{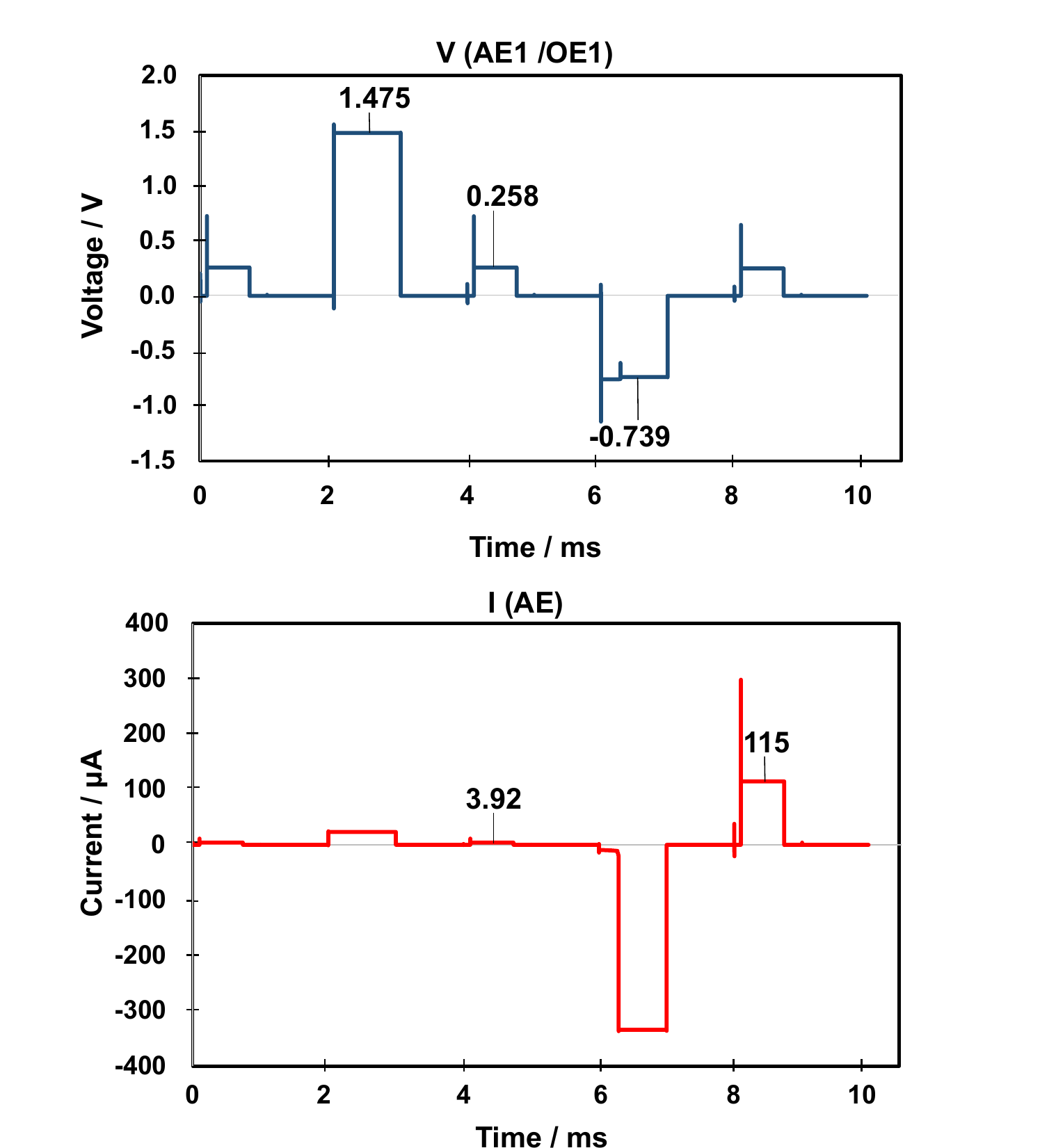} 
    \caption{Transient analysis using read verify algorithm with layout parasitics}
    \label{fig:enter-label}
\end{figure}

\subsection{Current Mode Operation}
In this mode, a regulated current is passed through the memristor through a specific time interval, which changes the conductance of the memristor as per Fig. 4(a).
 The amplitude of current corresponds to the DAC input value, $ \textit{V}_\text{ref} $, and the resistance $\textit{R}$ is,

\begin{equation}
    I_\text{mem} = \frac{V_\text{dd} - V_\text{ref}}{R} .
\end{equation}

The circuit functions as a voltage-to-current converter. A change in reference voltage results in a change in regulated current value. The conductance change is confirmed by evaluating voltage across the memristor, with constant current through it. A parametric analysis with \textit{V}\textsubscript{ref} ranging from \SI{300}{\milli\volt} to \SI{1.6}{\volt}, which results in regulated current ranging from \SI{3.5}{\micro\ampere} to \SI{20}{\micro\ampere} is depicted in Fig. 6.

\begin{figure}[htbp]
    \centering
    {\includegraphics[width=1\linewidth]{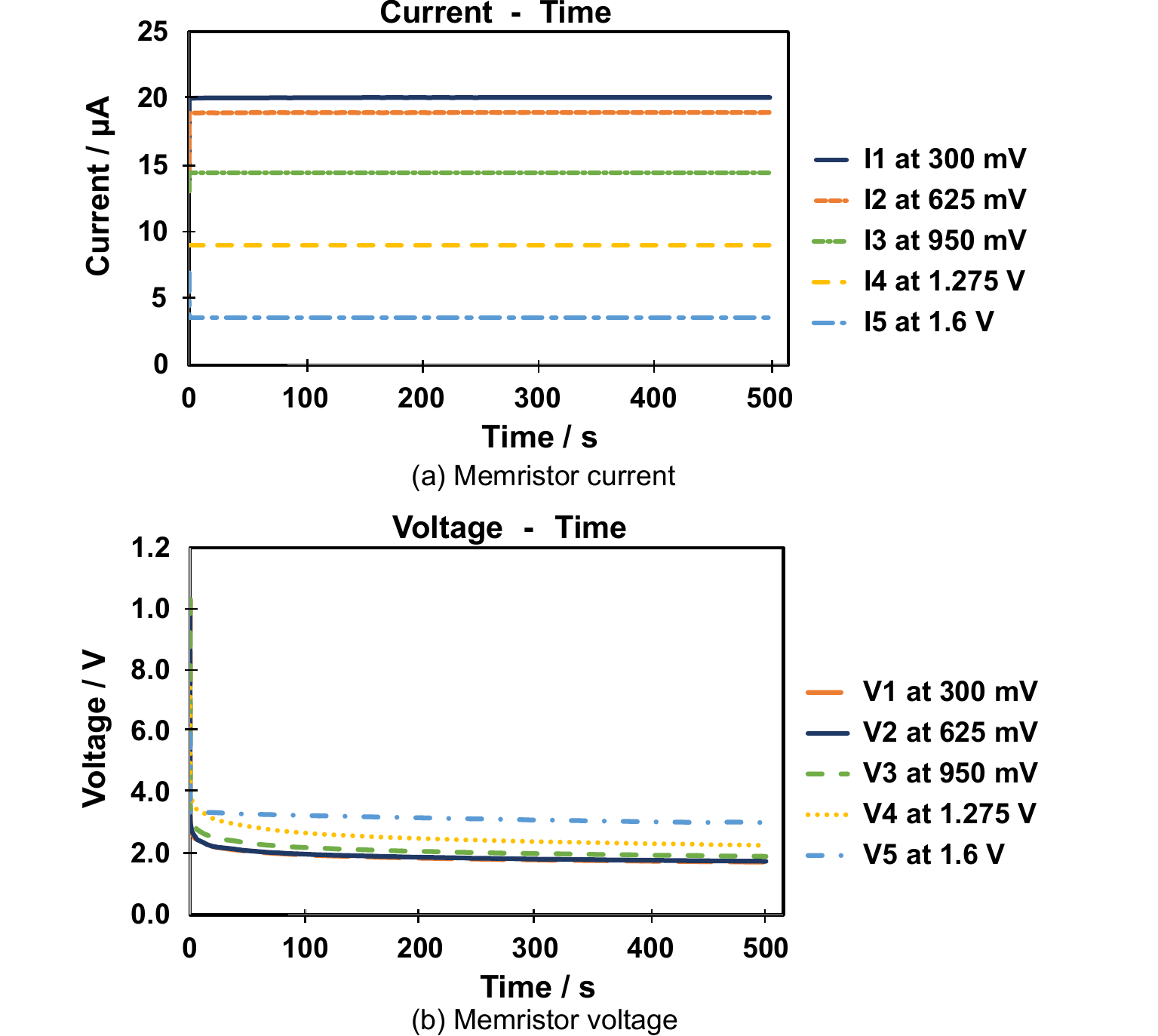}}
    \caption{Current through memristor (a) and Voltage across memristor (b), for reference voltage varying from \SI{300}{\milli\volt} to \SI{1.6}{\volt}}
    \label{fig:enter-label}
\end{figure}

The current mode operation differs from the voltage mode counterpart in the case of low resistance switching only. The memristor should be switched to HRS by the voltage mode mechanism before performing the write operation in LRS mode.
\subsection{Read Operation}
Unlike in traditional memristor based crossbars where a TIA is used to sum the column currents, here a current mode SAR ADC is employed to source the column currents, and acts as an active supply with regulated read voltage up to \SI{700}{\milli\volt} as per Fig. 4(b). To avoid affecting the memristor conductance state, the read voltage range is restricted to \SI{250}{\milli\volt}. The reference voltage for regulation is given to the input of the opamp and mimics the RESET state except that the supply is switched to the ADC.

\subsection{Analog Conductance}
An Analog conductance, i.e the multiple conductance states in the memristor, can be achieved by adjusting the amplitude and frequency of the voltage pulses applied across the memristors in voltage mode operation and by adjusting the current through the memristors with respect to time in current mode. In the current mode, the resistance is switched from HRS to LRS at a  faster rate, if there is an increase in the supply current to the memristor as shown in Fig. 7. The conductance increases exponentially with respect to current applied and time and can be depicted as:
\begin{equation}
    G_\text{mem} \propto I_\text{mem}\times e^{\text{t}} .
\end{equation}

\begin{figure}[h]
    \centering
    {\includegraphics[width=1\linewidth]{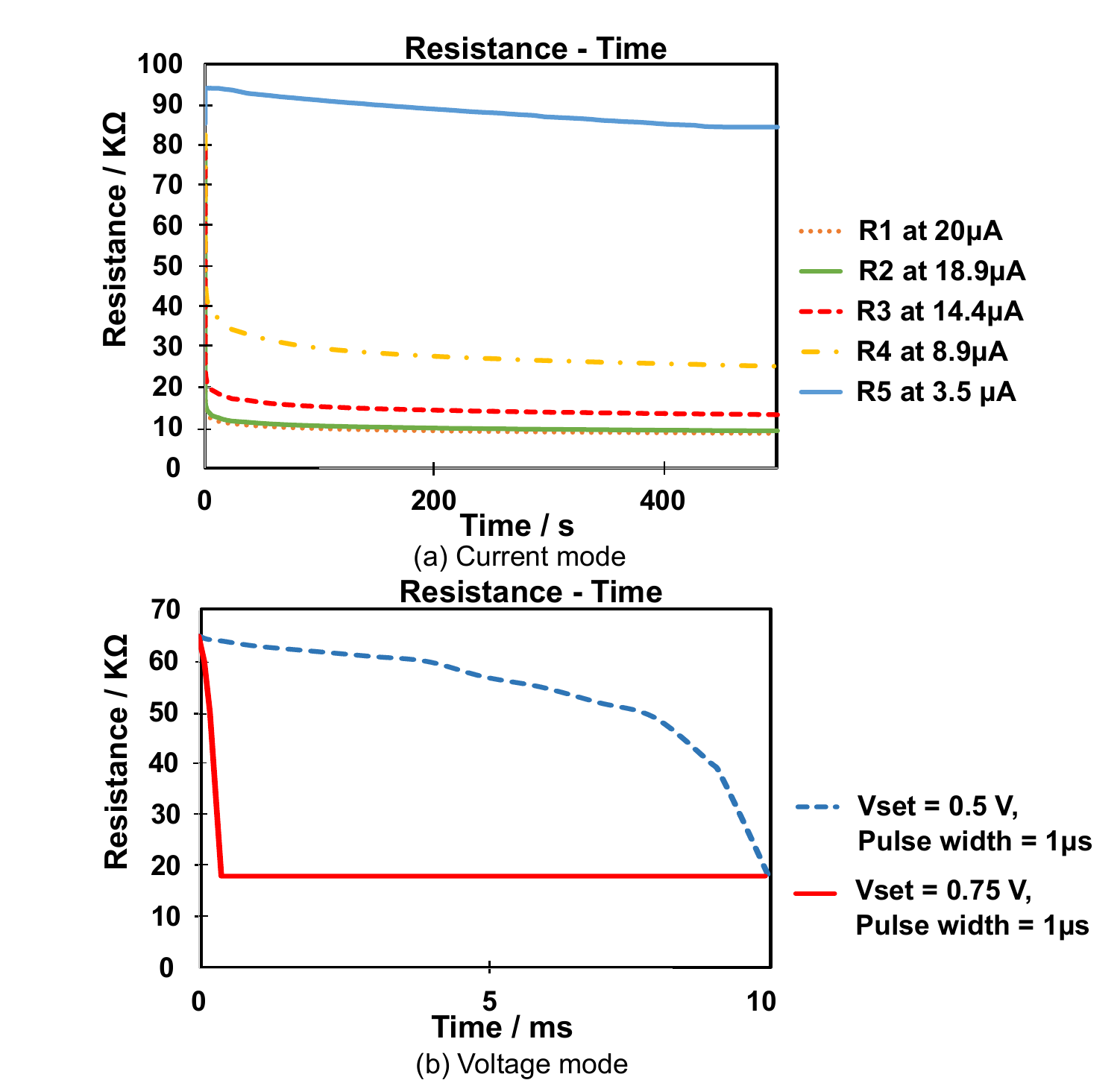}}\hfill
      \caption{Analog conductance in current mode (a) and voltage mode (b)}
    \label{fig:enter-label}
\end{figure}

\section{2T1R Crossbar Array Architecture}
A crossbar architecture that incorporates voltage mode and current mode is developed, where conductance control and VMM operation could be performed as shown in Fig. 8. If the voltage is represented as a vector and the conductance as a matrix, as per Kirchhoff's law, the vector matrix multiplication results in a current\cite{b9}.
\begin{figure}[htbp]
    \centering
    \includegraphics[width=1\linewidth]{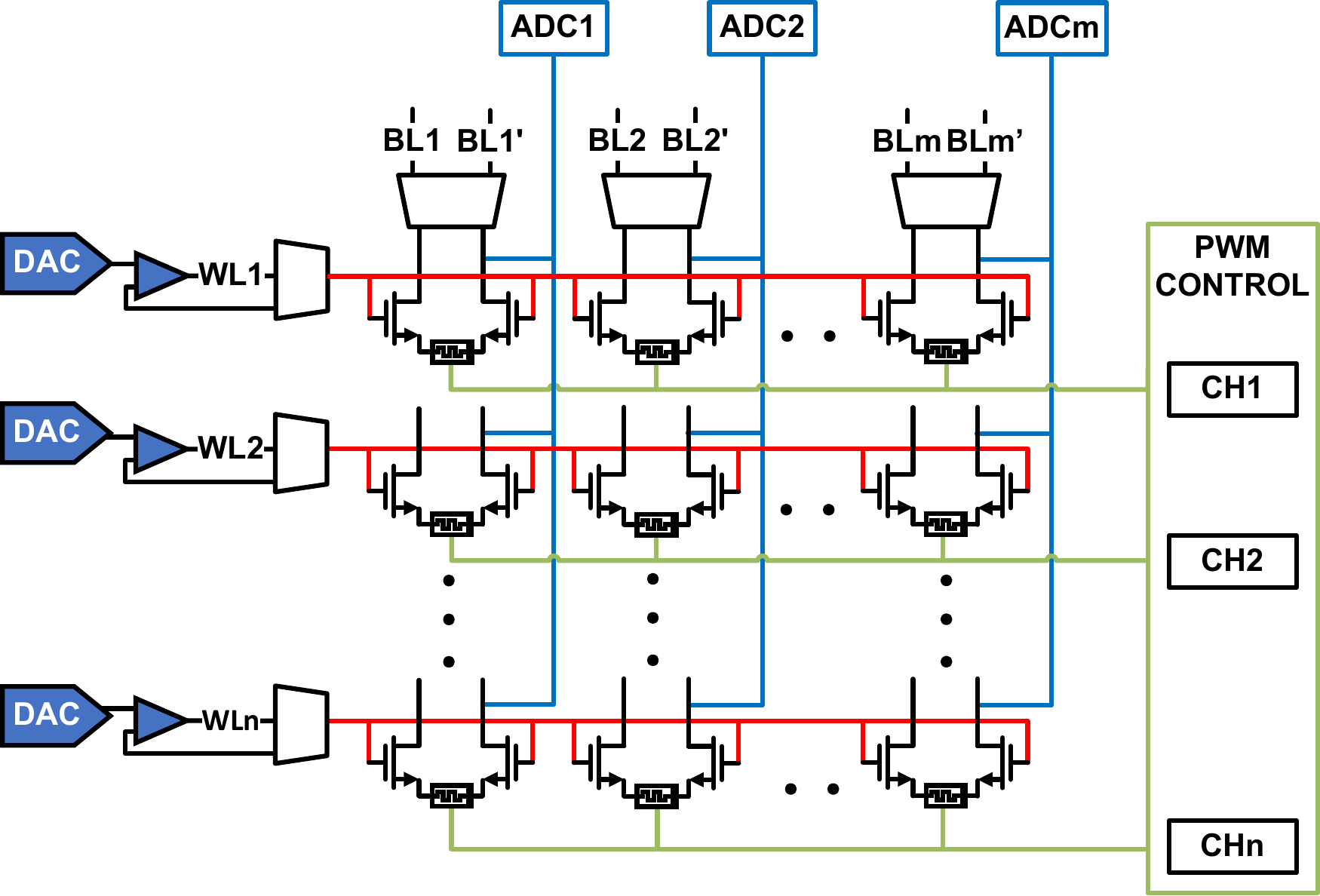 }
    
    \caption{2T1R regulated memristor conductance control array architecture}
    \label{fig:enter-label}
\end{figure}

The voltages generated by each DAC can be represented by $\textit{V} \in \mathbb{R}^{1 \times n}$. The programmed conductance matrix and the resulting current can be represented by $[\textit{G}] \in \mathbb{R}^{n \times m}$ and $[\textit{I}] \in \mathbb{R}^{1 \times m}$, respectively.

The memristor crossbars perform VMM operation in parallel compared to von Neumann computation, where the multiply accumulate operation is done with many sequential memory accesses. The bit lines correspond to either $\textit{V}_\text{dd} $ or ground. The analog conductance can be programmed (via word lines) to the matrix using either voltage mode or current mode before the application of the intended voltage matrix. The resulting current, which is added up in a column, could be sensed by a current mode ADC.
The architecture with an array size of 2x2 inclusive of regulated voltage mode and current mode, was simulated. The selection of rows and columns for different modes of operations as well as read and write operations were done using MUX selectors. Let the DAC inputs vector be,
\begin{equation}
V = \begin{bmatrix}\SI{0.25}{\volt}&\SI{0.25}{\volt}\end{bmatrix}
\end{equation}
and the resistance matrix be:
\begin{equation}
   R = \begin{bmatrix}\SI{5}{\kilo\Omega}&\SI{1.8}{\kilo\Omega}\\
   \SI{3}{\kilo\Omega}&\SI{65}{\kilo\Omega}\\
   \end{bmatrix} 
\end{equation}
Then, the currents obtained via column 1 and column 2, after simulations with parasitics are:
\begin{equation}
  I = \begin{bmatrix}\SI{129}{\micro\ampere}&\SI{135}{\micro\ampere}\end{bmatrix}
\end{equation}

The PWM is programmed to provide a pulse width of \SI{1}{\milli\second} and the simulation is performed for \SI{10}{\milli\second}. The resulting current through individual memristors and the total current through two columns are depicted in Fig. 9. Currents I1 and I3 correspond to column 1 current. Similarly, current I2 and I4 correspond to column 2 current.
\begin{figure}[h]
    \centering
    \includegraphics[width=1\linewidth]{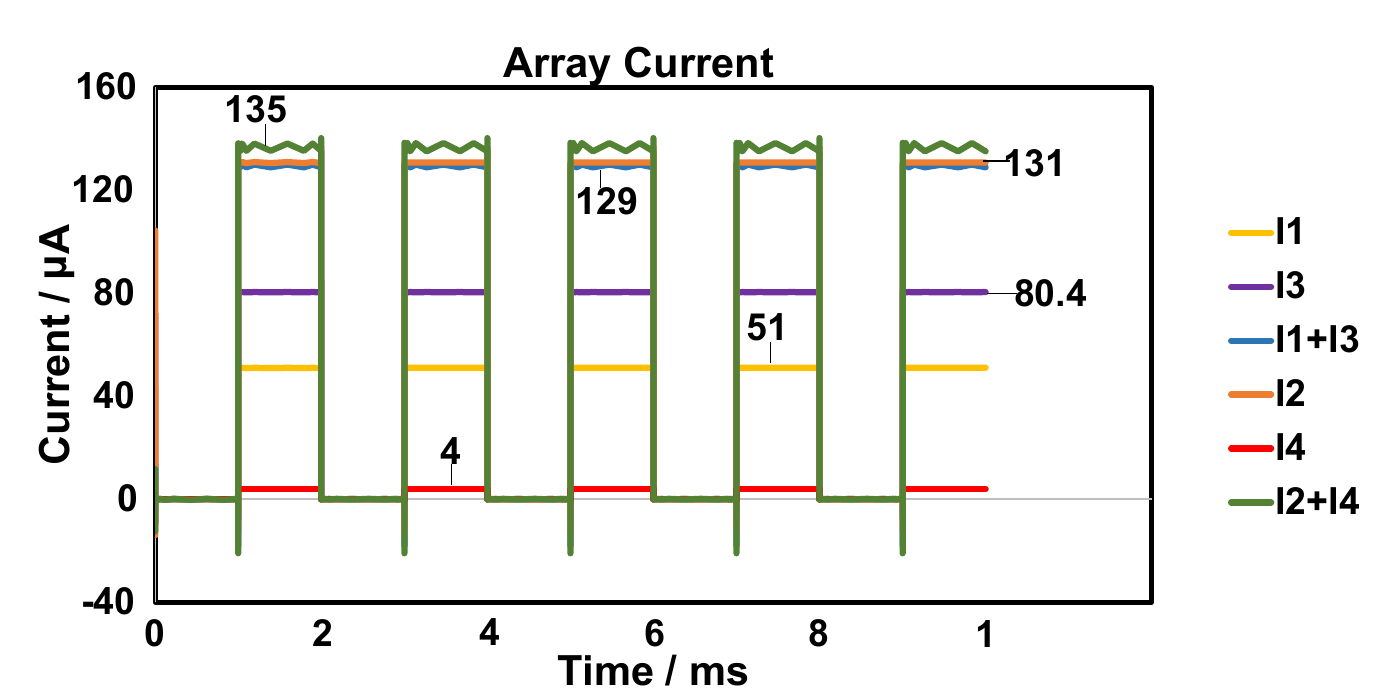}
    \caption{Current in 2x2 array architecture including parasitic effects}
    \label{fig:enter-label}
\end{figure}

\section{CMOS Implementation}
 The control architecture with a 2x2 array size is successfully taped out in a \SI{28}{\nano\meter} technology. The MEMCTRL block includes a core that corresponds to the intended conductance control architecture, incorporating voltage mode control and current mode control in 2 rows, as well as core biasing circuits, resistive DAC, and a digital control block for the pulse width generator as shown in Fig. 10. The pulse width control and digital configurations for the selection of rows and columns are done with the RISC V processor within, running with a clock frequency upto \SI{100}{\mega\hertz}. Alternatively, direct access is enabled through a JTAG programming interface. The readout circuit is implemented with the current source SAR ADC. The chip layout for the same is shown in Fig. 10. The overall size of the chip is \SI{1.4}{\milli\meter} by \SI{1.0}{\milli\meter}, with which the memristor control circuit incorporates an area of \SI{0.4}{\milli\meter} by \SI{0.5}{\milli\meter}. The memristors are not co-integrated in this architecture, however, external pins are provided for accessing external memristors.
 
\begin{figure}[h]
    \centering
    \includegraphics[width=1\linewidth]{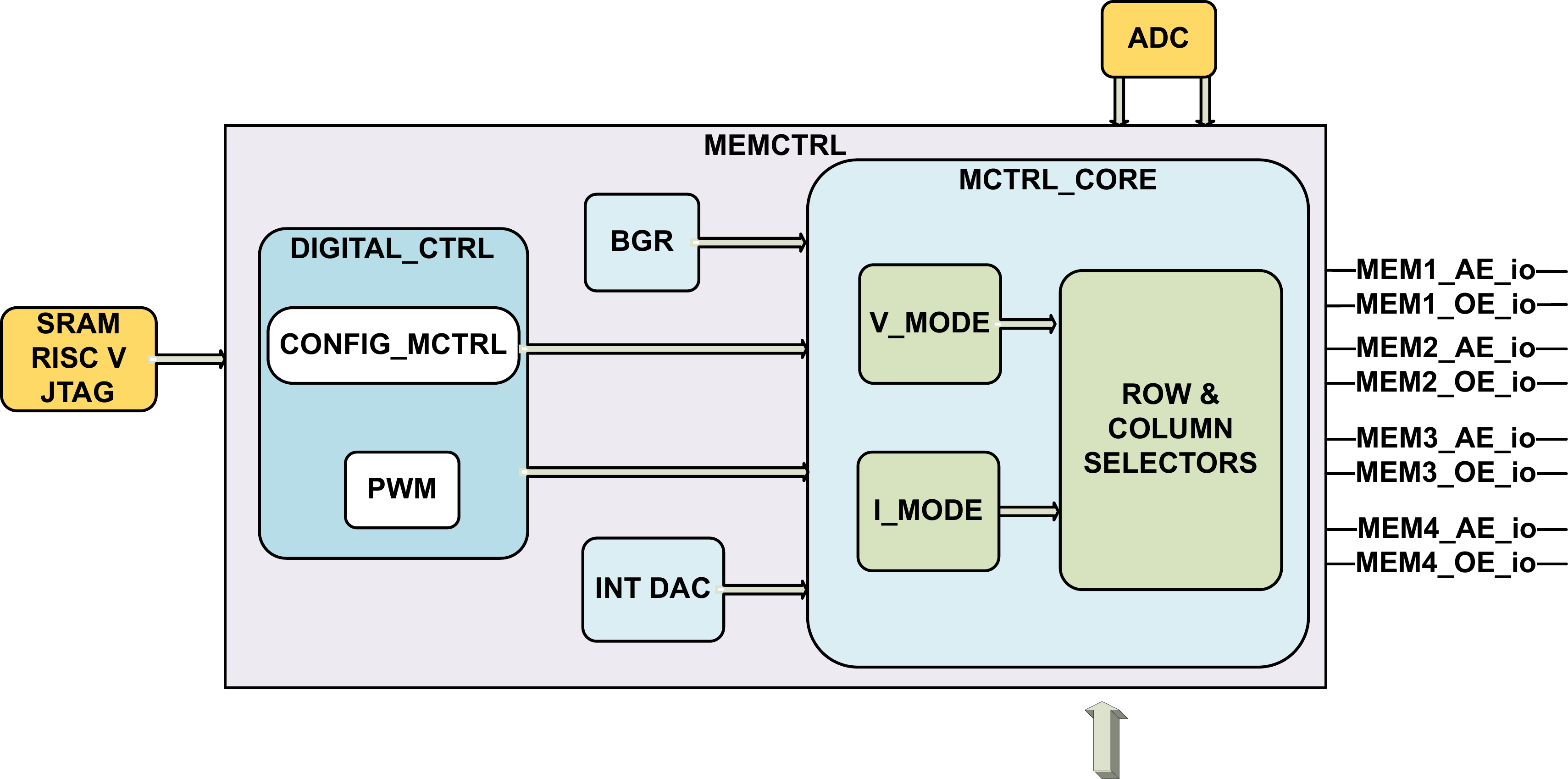}\\
    \includegraphics[width=0.8\linewidth]{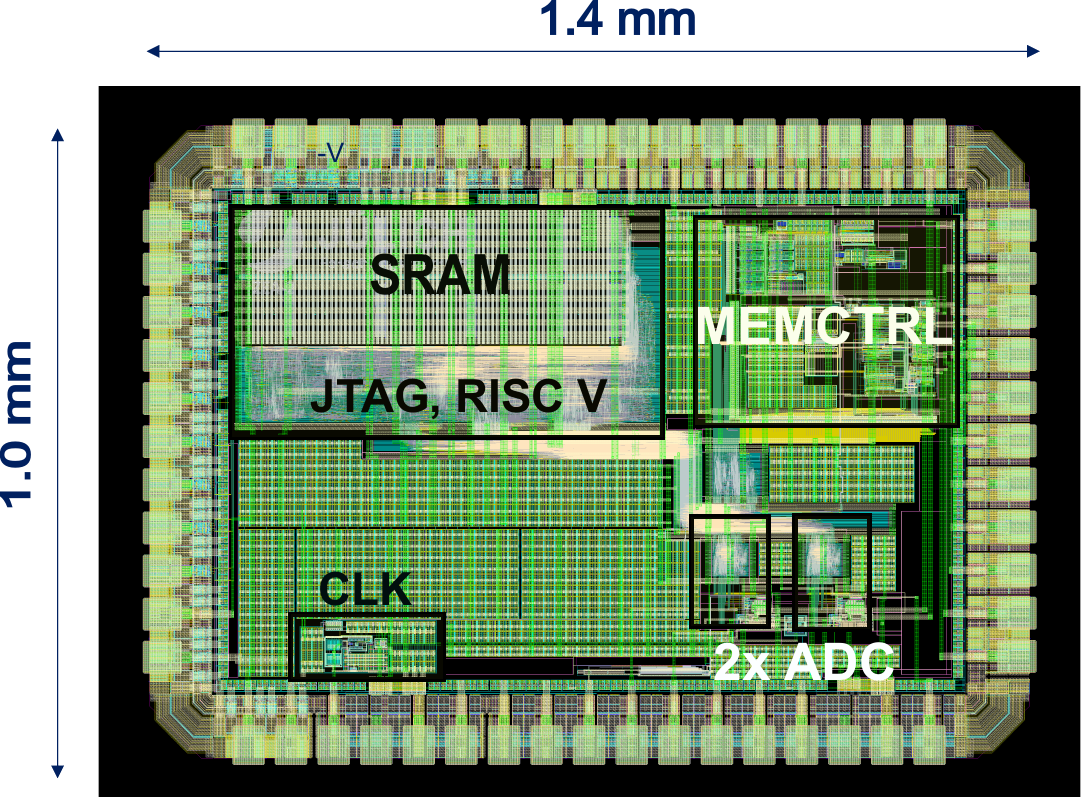}
    \caption{Chip layout of the manufactured memristor control demonstrator with annotated blocks}
    \label{fig:enter-label}
\end{figure}

In the conductance control core circuit, the maximum power drawn is \SI{717.1}{\micro\watt}. The power drawn by individual devices in different modes is shown in Table III. The opamp is the prominent source for current consumption. As opamp is shared per row, the power consumption can be reduced significantly.

\begin{table}[htbp]
\caption{Power Analysis}
\begin{center}
\begin{tabular}{|c|c|c|}
\hline
\textbf{ }&\multicolumn{2}{|c|}{\textbf{Device Number}} \\
\cline{2-3} 
\textbf{Mode} & \textbf{\textit{1 Device}}& \textbf{\textit{2x2 Devices}} \\
\hline

Voltage control&  \SI{412.1}{\micro\watt} & \SI{717.1}{\micro\watt} \\
\hline
Current control& \SI{349.4}{\micro\watt} & \SI{556.03}{\micro\watt} \\
\hline
\end{tabular}
\label{tab1}
\end{center}
\end{table}
\section{Conclusion}
A 2T1R regulated memristor conductance control array architecture is implemented and taped out in \SI{28}{\nano\meter} CMOS technology. Two modes of analog conductance control are designed and simulated with the given architecture. The amplitude and time period of the applied input source has a significant impact on the conductance of the memristor state. The current mode could be considered as an alternative technique to the voltage mode for conductance control. One opamp per row is required for regulation in voltage mode and subsequent current mode. Traditional TIA-based readout is replaced by current mode SAR ADC, where current sourcing rather than sinking is done during the readout process. A sneak path current is avoided in our architecture as both terminals of memristors are grounded when not selected. The conductance control operation via memristor selection can be achieved by programming the analog switches in between through the RISC-V core. The array architecture is developed to study the impact of regulated signals on the conductance of memristor with multiple select lines for different modes of operation and could be optimized in the future, with better programmability conditions.

\section{Acknowledgement}
The memristor model was provided by the Peter Grünberg Institute (PGI)-7, Forschungszentrum, J\"ulich. Especially, we want to thank Dr.-Ing. Christopher Bengel, for his support in the memristor model. This research work is a part of the work package of the demonstrator project of NEUROTEC II (https://www.neurotec.org), carried out at Forschungszentrum J\"ulich and funded by the German Federal Ministry of Education and Research (BMBF grant number 16ME0398K).

\begin{equation}
 V = \begin{bmatrix}v_1&v_2&\hdots&v_n\end{bmatrix} 
\end{equation}

The conductance matrix is represented by 
\begin{equation}
  G = \begin{bmatrix}g_{11}&g_{12}&\hdots&g_{1m}\\
  g_{21}&g_{22}&\hdots&g_{2m}\\
   \vdots&\vdots&\ddots&\vdots\\
   g_{n1}&g_{n2}&\hdots&g_{nm}
   \end{bmatrix} 
\end{equation}

The vector matrix multiplication can be represented as
\begin{equation}
 I = \begin{bmatrix}v_1&v_2&\hdots&v_n\end{bmatrix}.\begin{bmatrix}
  g_{21}&g_{22}&\hdots&g_{2m}\\
  \vdots&\vdots&\ddots&\vdots\\
   g_{n1}&g_{n2}&\hdots&g_{nm}
   \end{bmatrix}
\end{equation}
\begin{equation}
  I = \begin{bmatrix}\sum_{i=1}^nv_ig_{i1}&\sum_{i=1}^n v_ig_{i2}&\hdots&\sum_{i=1}^n v_ig_{im}\end{bmatrix}
\end{equation}


\begin{thebibliography}{00}
\bibitem{b1} H. Bao et al. ``Toward memristive in-memory computing: principles and applications,'' Frontiers of Optoelectronics, vol. 15, Issue. 1, December 2022, doi: 10.1007/s12200-022-00025-4


\bibitem{b2} K. Kim et al. ``A Functional Hybrid Memristor Crossbar-Array/CMOS System for Data Storage and Neuromorphic Applications'', vol. 12, Issue. 1, pp. 389--395, January 2012, doi: 10.1021/nl203687n

\bibitem{b3} A. N. Youssef, A. L. Jagath, N. K. Thulasiraman, H. A. Almurib, " Effect of Sneak Path Current in TiOx/HfOx Based 1S1R RRAM Crossbar Memory Array", IEEE SCOReD, November 2021, pp. 267-272, doi: 10.1109/SCOReD53546.2021.9652781

\bibitem{b4} L. Shi, G. Zheng, B. Tian, B. Dkhil and C. Duan, ``Research progress on solutions to the sneak path issue in memristor crossbar arrays'', Nanoscale Adv., vol. 2, Issue. 5, pp. 1811--1827, 2020, doi: 10.1039/D0NA00100G.

\bibitem{b5} W. L. Ahmad, D. Wouters, C. Bengel, R. Waser, S. Menzel, ``Analysis of VMM Operations on 1S1R Crossbar Arrays and the Influence of Wire Resistances'', IEEE ISCAS, May 2022, pp. 91-95, doi: 10.1109/ISCAS48785.2022.9937570

\bibitem{b6} Rainer Waser(Ed.), Nanoelectronics and Information Technology, Wiley-VCH, ISBN-13: 978-3-527-40927-3

\bibitem{b7} C.Bengel et al. ``Variability-Aware Modeling of Filamentary Oxide-Based Bipolar Resistive Switching Cells Using SPICE Level Compact Models'',IEEE Trans. Circuits Syst. I, vol. 67, Issue. 12, December 2020, pp. 4618-4630, doi: 10.1109/TCSI.2020.3018502 

\bibitem{b8} R. Jacob Baker, Harry W. Li, David E. Boyce, CMOS Circuit Design, Layout, And Simulations, Wiley-IEEE Press ISBN 0-7803-3416-7
\bibitem{b9} J. Chen, J. Li, Y. Li, X. Miao, ``Multiply accumulate operations in memristor crossbar arrays for analog computing'', J. Semicond., vol. 42, Issue. 1, pp. 013104, January 2021, doi: 10.1088/1674-4926/42/1/013104

\end{thebibliography}
\end{document}